\documentclass[pre, reprint,superscriptaddress, showkeys]{revtex4-2}
\usepackage[utf8]{inputenc}
\usepackage[english]{babel}
\usepackage{amsmath}
\usepackage{amsfonts}
\usepackage{amssymb}
\usepackage{graphicx}
\usepackage{siunitx}
\usepackage{tabularx}
\usepackage{booktabs}
\usepackage{multirow}
\usepackage{hyperref}
\begin{document}
\title{Characterization of the gelation and resulting network of a mixed-protein gel
derived from sodium caseinate and ovalbumin in the presence of glucono-$\delta$-lactone}
\author{Naoko Yuno-Ohta}
\email[Corresponding Author:]{NaokoOHTA@aol.com}
\author{Naoya Shimonomura}
\author{Yuuka Hoshi}
\affiliation{Department of Food and Nutrition Advanced Course of Food and Nutrition\\
Junior College at Mishima, Nihon University, 2-31-145 Bunkyo-Cho Mishima City Shizuoka pref., Japan}

\author{Mathieu Leocmach}
\affiliation{Université de Lyon, Université Claude Bernard Lyon 1, CNRS, Institut
Lumière Matière, F-69622, VILLEURBANNE, France}

\author{Koichi Hori}
\author{Hiroyuki Ohta}
\affiliation{School of Life Science and Technology, Tokyo Institute of Technology,
4259-B65 Nagatsuta Yokohama Kanagawa pref., Japan}

\begin{abstract}
We investigated mixed-protein gels made from sodium caseinate and ovalbumin at different ratios with use of the acidification agent glucono-$\delta$-lactone. Dynamic viscoelastic measurements revealed that increasing the ovalbumin content decreased the mechanical properties of the gel but accelerated onset time of the phase transition. Ultrasound spectroscopy during gelation revealed that the relative velocity gradually decreased, whereas the ultrasonic attenuation increased during the whole acidification process until gelation was complete, although these changes were much smaller than those observed with heat-induced gelation. Confocal laser scanning microscopy along with scanning electron microscopy revealed that although uniform mixing of sodium caseinate and ovalbumin was observed, sodium caseinate is likely to mainly lead formation of the gel network, and the porosity of the resulting gel network depends on the ratio of these two components. The results demonstrate that confocal laser scanning microscopy is a useful tool for analyzing both the networks within mixed-protein gels and the contribution of each protein to the network and gelation.
\end{abstract}

\maketitle

\section{Introduction}

The interactions between different proteins in mixed-protein
gel systems are interesting because they are able to produce novel
physicochemical properties of gels as shown in several reports \cite{matsudomiACaseinImprovesGel2003,messionEffectGlobularPea2015}.
However, some details remain unknown—for example, how an individual
protein contributes to the final gel network.

In the past two decades, food scientists have adopted ultrasound
spectroscopy as a standard method for analysis of gelation mechanisms of
food constituents. For example, Povey et al.~\cite{poveyInvestigationBovineSerum2011} investigated the ultrasonic
properties of mildly and severely denatured bovine serum albumin,
namely, both the velocity and attenuation changes of ultrasound during
the process were monitored. They found that mild heat treatment caused
the albumin molecules to dimerize or trimerize without any change in
velocity of sound-wave propagation, whereas severe heat treatment
resulted in irreversible denaturation and gelation, which affected both
ultrasound velocity and attenuation.
We previously investigated the effects of sodium caprate, (a type of
fatty-acid sodium salt) on the formation of mixed-protein gels of
ovalbumin
ultrasound spectroscopy, Fourier transform–infrared spectroscopy, and
scanning electron microscopy (SEM)~\cite{yuno-ohtaEffectsAcaseinSodium2014}. In the presence or absence of
sodium caprate, heating the $\alpha$-casein–OVA mixture resulted in a
transparent or translucent gel.
(OVA)
and
$\alpha$-casein
using
rheological
measurements,
Furthermore, in other work, we reported that $\beta$-casein contributes to sodium caprate–induced gelation with one
type of heat-coagulable protein, namely milk whey protein, without
heating~\cite{yuno-ohtaVCaseinAidsFormation2011}. These studies suggested that room-temperature, slow gelation
facilitates the analysis of soluble aggregates during gelation. Ultrasound
spectroscopy is a useful technique for studying the formation of soluble
aggregates prior to the phase transition, i.e., from soluble proteins to gels.
Using a mixed-protein system comprising casein and whey proteins,
Vasbinder A. J. et al.~\cite{vasbinderGelationCaseinWheyProtein2004} found that the final network structure differed
depending on whether the whey protein was preheated or subjected to
some other pretreatment. Their results suggested that mixing denatured
whey protein with casein results in a specific three-dimensional gel
network. Lopez et al.~\cite{lopezDiffusingWaveUltrasonic2009} investigated rennet-induced gelation of milk
protein in the presence of high-methoxyl pectin. They observed that
increasing the pectin content altered the microstructure of the gel and
that skim milk containing no pectin (control) or only 0.04\% pectin (in the
presence of pectin at low concentration) could form a rennet-induced gel.
The gel had a fine microstructure without pectin, whereas a more open
network formed in the presence of pectin. They reported that confocal
laser scanning microscopy (CLSM) techniques could be used to monitor
the evolution of the network. However, their use of a single fluorescent dye
physically absorbed in the network gave no detailed information on the
composition (protein and polysaccharide mixture) of the resulting gel.
Here, we studied the acid-induced gelation of a mixture of sodium
caseinate (SC) and OVA at room temperature. Actually, their sources,namely milk and egg, are often used with flour as a dough for making
flour-paste products such as hot cakes or doughnuts, and it is therefore of
interest to know the interactions between both proteins. By adding a
specific fluorescent probe to each of SC and OVA prior to mixing the
proteins, we could visualize interactions between soluble aggregates using
CLSM. We also performed several instrumental analyses (ultrasound
spectroscopy, rheological measurement and SEM) with imaging analysis
by CLSM. The addition of a different fluorescent probe to each of SC and
OVA prior to mixing the proteins facilitated the visualization of
interactions between the soluble aggregates using CLSM. The results
were reinforced with several other instrumental analyses, namely
ultrasound spectroscopy, rheological measurements, and SEM.

\section{Materials and Methods}
\subsection{Preparation of samples}
Commercial SC (Alanate 180 made in New Zealand Milk Protein, Fontera
Japan) was donated by Meiji Co. Ltd. (Japan) and used without further
purification. OVA (A-5378 Grade III) was purchased from Sigma-Aldrich
(USA), and glucono-$\delta$-lactone (GDL) was from Fujifilm Wako (Japan).
Prior to the instrumental analysis, a solution of 10\% (w/v) SC or OVA was
prepared in distilled water containing 0.025\% NaN$_3$ . For the mixed-protein
system, the two protein solutions were mixed at two different ratios:
SC:OVA = 8:2 or 5:5. Finally, GDL powder (2.5\%) was gradually added to
the two-protein mixed solution with continuous mixing using a touch mixer.The
pH
of
the
solution
was
monitored
continuously
with
a
microprobe-equipped hand-held pH meter. Protein concentration was
usually determined by gravimetric analysis except for the protein sample of
CLSM (see 2.4 CLSM).

\subsection{Rheological measurements}
Dynamic modulus parameters (storage modulus ($G^{\prime}$), loss modulus ($G^{\prime\prime}$),
complex modulus ($G^{*}$) and loss tangent ($\tan\delta$) were determined by dynamic
viscoelastic measurement (Rheolograph Sol, Toyo-seiki Ltd., Tokyo, Japan).
The sample solution was placed between parallel plates, and the gap
between the blade and both sides of the plates was set to 1 mm. The sample
solution (1.6 ml) was subjected to longitudinal
(displacement magnitude $\pm$\SI{50}{\micro\metre}) of 1 Hz frequency at constant
temperature (\SI{25}{\celsius})~\cite{nishinariNewApparatusRapid1980,yuno-ohtaFormationFattyAcid1998}. At least three independent experiments were
carried out for each sample. Data were analyzed and plotted using
Microsoft Excel.
shear oscillation

\subsection{Ultrasonic spectroscopy}
Ultrasonic spectroscopy was performed using an HR-US 102 instrument
(Ultrasonic Scientific, Dublin, Ireland) provided by ST Japan (Tokyo,
Japan). The data were processed with the manufacturer’s software (version
5.43). The spectrometer generates transverse sound waves that pass
through both the sample and reference cells and measures both the velocity
and attenuation of the transmitted sound wave. Solutions were prepared as described in 2.1. Sample solutions were degassed using a centrifuge (50 g ,
3 min, ambient temperature of $\approx$\SI{25}{\celsius}) after adding GDL and equilibrated
at \SI{25}{\celsius}. After loading 1 ml of each of the sample and reference (water)
solutions into their respective cells, changes in both the velocity and
attenuation of the sound waves at a selected frequency (2.5, 5, and 8 MHz)
were continuously monitored in both the sample and reference cells with
time. Prior to the experiments, the frequency was set to 2.5, 5, and 8 MHz,
and calibrated with degassed water at \SI{25}{\celsius}. The velocity of sound waves
through each of the sample and reference cells was recorded separately.
The internal temperature of the solution was controlled via a circulator
(Haake Phoenix II) with a bath (Haake C25P) (Thermo Fisher Scientific,
Newington, NH, USA). At least three independent experiments were
carried out for each sample. Data were analyzed and plotted using
Microsoft Excel.

\subsection{CLSM}
The dyes used were amino-reactive N-hydroxysuccinimide esters that
react with primary amines in proteins, resulting in a covalent amide bond
and release of the N-hydroxysuccinimide group. From Thermo Scientific,
we purchased two fluorescent dyes having different excitation and emission
wavelengths: DyLight 405 (Ex/Em: 400 nm/420 nm) for SC, and DyLight
633 (Ex/Em: 638 nm/658 nm) for OVA. The labeling was performed
according to the instruction manual provided by the company with a slight
modification, namely, we dissolved each dye (\SI{50}{\micro\gram}) in dimethyl formamide
(\SI{50}{\micro\litre}) and then mixed it with the protein ($\approx$15\% by weight) in 10 ml sodium
borate, pH 8.5. The mixture was incubated at room temperature for 1 h and
then dialyzed in the sodium borate buffer to clear any unreacted dye. The
protein concentration was determined by the Lowry method~\cite{lowry1951protein}. When
needed (less than 10\%), each protein solution was concentrated using a
centrifuge tube with a nonionic membrane (nonionic serve 010 type) which
produced from GE healthcare life science. To prepare each sample, we
mixed dye-treated SC and OVA solutions (total protein concentration 10\%)
with GDL powder (final concentration 2.5\% by weight and at ambient
temperature around \SI{25}{\celsius}) and immediately placed the mixture (\SI{50}{\micro\litre}) into
the gap between the cover and slide glass using parafilm (thickness $\approx$\SI{100}{\micro\metre}), sealed the edges of the cover glass with manicure sealant, and
monitored the pH change over time.
CLSM images were obtained using a LSM 780 confocal microscope (ZEISS) alternating between 488nm and 543nm exitations at ambient temperature ($\approx$\SI{25}{\celsius}) with the following settings:
DyLight 405 (excitation, 405 nm with a diode laser; detection range,
409–552 nm) and DyLight 633 (excitation, 633 nm with a He-Ne laser;
detection range, 637–746 nm).
Images were processed and analyzed with Zeiss ZEN 2012 software.	

\subsection{SEM}
The gels were formed during acidification by GDL (2.5\%) at ambient
temperature ($\approx$\SI{25}{\celsius}). Aliquots were taken at different pH values and rinsed
immediately with 0.5\% (w/v) glutaraldehyde to halt any further change in
pH of the gel. Then, each gel sample was followed by chemical fixation
using 0.5\% osmium tetroxide in 0.1 M sodium phosphate buffer (pH 7), and
then conductive staining was performed using 0.5\% (w/v) filtered tannic
acid. This double-fixation method followed by conductive treatment with
0.5\% tannic acid was repeated according to methods by Yuno-Ohta et al.~\cite{yuno-ohtaRoleOvomucoidGelation2016}
to prevent solubilization of gels and ensure fixation. Gels were then
dehydrated in a series of ethanol solutions of increasing concentration (50,
70, 80, 90, 95 and 100\% and again 100\%, v/v). Following dehydration,
ethanol was completely replaced with tert-butyl alcohol by soaking in
tert-butyl alcohol solutions of increasing concentration (20, 40, 70, and
100\% and
again
in
100\%,
v/v).
Samples
were
dried
using
a
tert-butyl-alcohol dryer (Sinku Devices Co. Ltd., VFD-21S, Ibaraki, Japan)
and then glued to Au (150-Å thickness) on a quick auto coater SC-701
(Sanyu Denshi Co. Ltd.). Microscopy was performed with a SEM
(JSM-6010LA type, JEOL, Japan).

\section{Results}
\subsection{Rheology during gelation and the influence of OVA content}

\begin{figure}
\includegraphics[width=\columnwidth]{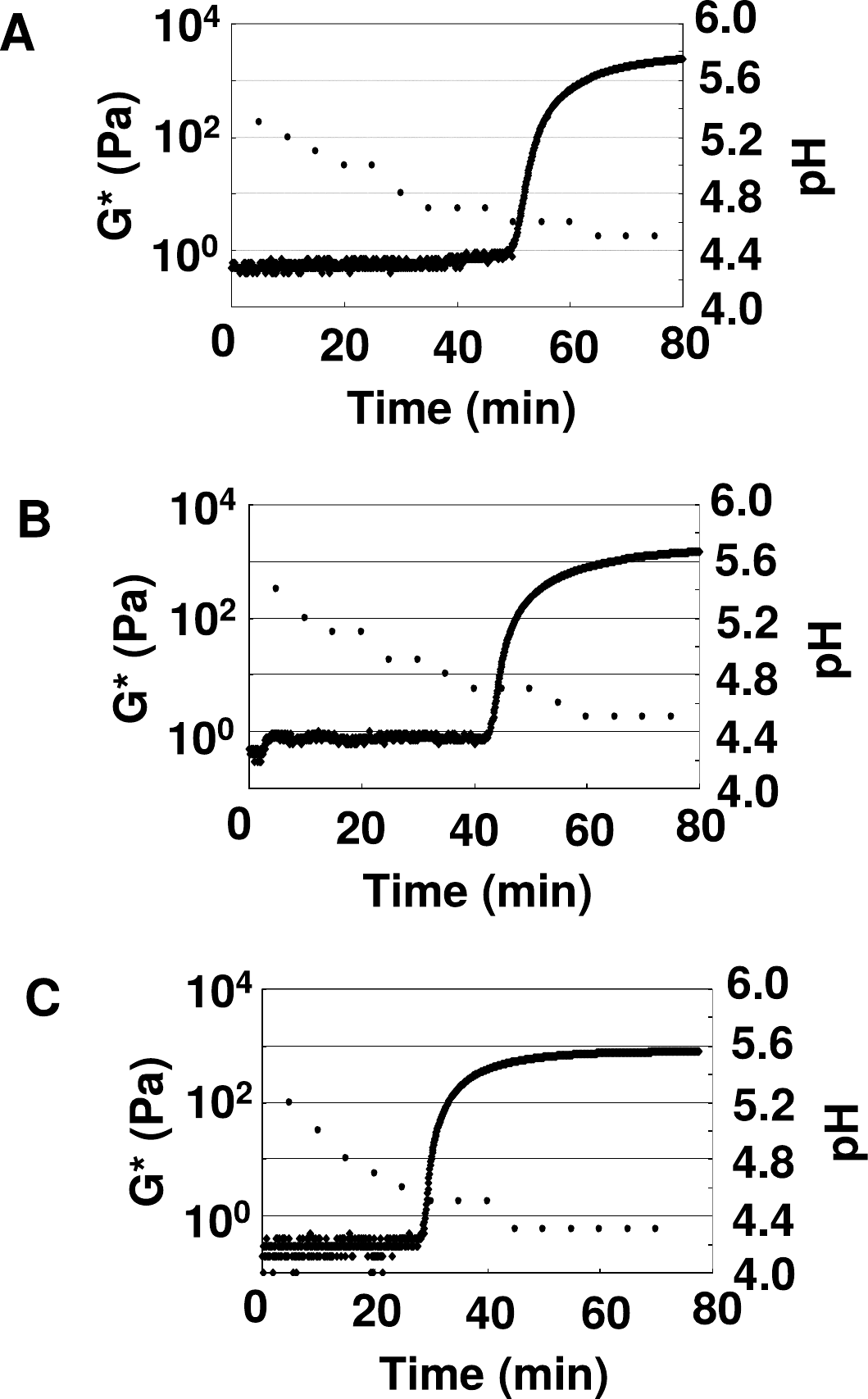}
\caption{Changes in dynamic complex modulus of SC with and without OVA.
A, 10\% SC; B, 8\% SC+2\% OVA; C, 5\% SC+5\% OVA.}\label{fig:rheology}
\end{figure}

Figure~\ref{fig:rheology} presents data concerning changes in the complex modulus ($G^{*}$)
and pH with time after GDL was added to the protein solutions. The
isoelectric point (pI) values for these proteins are similar, i.e., 4.44–4.76 for
$\alpha_{s1}$-casein, 4.83–5.07 for $\beta$-casein~\cite{eigelNomenclatureProteinsCow1984}, and 4.5–4.8 for OVA~\cite{powrie1986chemistry}, and thus
we expected that the response to a drop in pH should be independent of the
mixing ratio. Indeed, $G^{*}$ started to increase when the pH reached $\approx$4.5,
although the pH decreased more rapidly with higher OVA content.

Table~\ref{tbl:rheology} presents data for the $G^{\prime}$, $G^{\prime\prime}$, the gap between $G^{\prime}$ and $G^{\prime\prime}$, $\tan\delta$, $G^{*}$,
and the time at which the increase in complex modulus started. Generally,
the final value for $\tan\delta$ was smaller than 0.5, indicating that it is roughly
recognized as a gel. Furthermore, Ross-Murphy et al.~\cite{ross2012physical} reported when
the value of $G^{\prime}$ is >10-fold greater than that of $G^{\prime\prime}$, the gel is considered a
true gel~\cite{ross2012physical}. Therefore, the three protein systems we tested were
considered dense suspensions or weak gels~\cite{ikedaWeakGelType2001}. $G^{\prime}$ and $G^{\prime\prime}$ decreased as the
proportion of OVA increased. The dynamic modulus of the 5\% SC +5\% OVA
system increased most rapidly among the three gel formulations, perhaps
owing to the more rapid decrease of pH during gelation (Fig. S5).

\begin{table}
\begin{tabularx}{\linewidth}{@{}rcrrrrrc@{}}\toprule
\multicolumn{2}{c}{Composition}&
\multicolumn{5}{c}{Rheology at 75min (Pa)}&
\multirow{2}{7em}{\centering Starting time to increase of $G^*$} \\\cmidrule(r){1-2}\cmidrule(lr){3-7}
SC&OVA&
$G^\prime$&
$G^{\prime\prime}$&
$G^\prime-G^{\prime\prime}$&
$G^*$&
$\tan\delta$\\\cmidrule(r){1-1}\cmidrule(r){2-2}\cmidrule(r){3-3}\cmidrule(r){4-4}\cmidrule(r){5-5}\cmidrule(r){6-6}\cmidrule(){7-7}\cmidrule(l){8-8}
10\%& 0\% &1960& 748& 1212& 2099& 0.38& 51 min\\
8\%&2\%& 1320& 444& 876& 1383& 0.34& 43 min\\
5\%&5\%& 752& 240& 512& 789& 0.32& 30 min\\\bottomrule
\end{tabularx}
\caption{Storage modulus (G’), loss modulus (G”) at almost reached
plateau values (at 75 min), gap of $G^\prime$ and $G^{\prime\prime}$ at 75 min, $\tan\delta$,
complex modulus G* (at 75 min) and starting time to increase of
$G^*$ during acidification with GDL.}\label{tbl:rheology}
\end{table}

\subsection{Ultrasonic velocity and attenuation during gelation}
\begin{figure}
\includegraphics[width=\columnwidth]{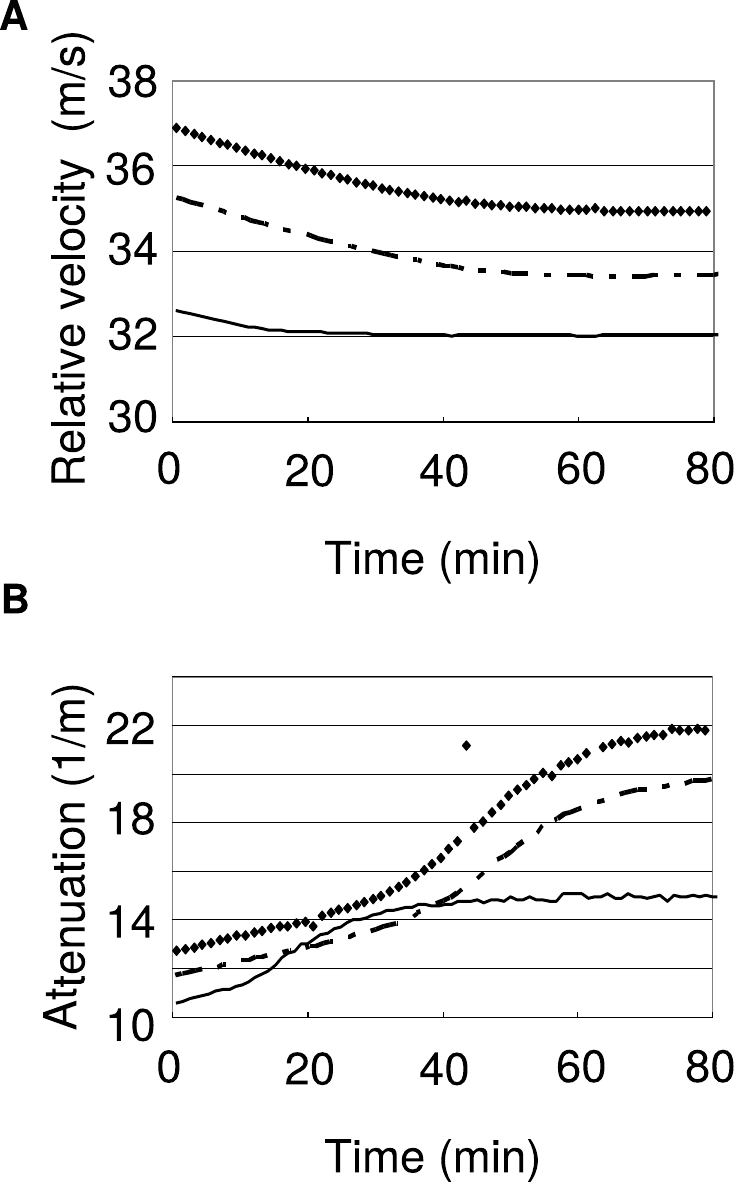}
\caption{Time dependence of the relative velocity (A) and attenuation (B)
of SC with without OVA. thick line, 10\% SC; dashed line, 8\%
SC+2\% OVA; thin line, 5\% SC+5\% OVA.}\label{fig:US}
\end{figure}
Figure~\ref{fig:US}A shows the change in ultrasonic velocity during GDL-induced
acidification. The velocity began to decrease immediately for all ratios of
mixed SC and OVA but plateaued sooner for the 5\% SC + 5\% OVA sample.
The decrease in velocity occurred before gelation began, whereas the
plateau in velocity corresponded to a gelled sample.
The change was approximately 1.8 m/s after 60 min for each of the 10\% SC
and the 8\% SC + 2\% OVA mixed systems, whereas the change was 0.6 m/s
for the 5\% SC + 5\% OVA system after 60 min, which initially decreased 0.6
m/s within the initial $\approx$20 min and remained constant thereafter. This
implied that the magnitude of the velocity decrease diminished with higher
OVA content. The extreme case was for 10\% OVA, for which velocity did not
decrease over the entire 80-min period (not shown), suggesting that SC is
the primary cause of the velocity reduction.

Figure~\ref{fig:US}B presents data for the change in attenuation, which increased
with time for all ratios but plateaued sooner for the 5\% SC + 5\% OVA
sample. The overall change in attenuation was approximately \SI{9}{\per\metre} for the
10\% SC system and the 8\% SC + 2\% OVA mixed system and \SI{4}{\per\metre} for the
5\% SC + 5\% OVA system.

The temporal velocity and attenuation changes were similar for the 10\%
SC sample and 8\% SC + 2\% OVA sample, whereas the increase in
attenuation of the 5\% SC + 5\% OVA sample was $\approx$60\% smaller than that of
the 10\% SC sample or 8\% SC + 2\% OVA sample and plateaued by 40 min.
258
We previously investigated the changes in ultrasonic velocity and
attenuation
during
the
formation of
a
sodium
caprate–induced
$\beta$-lactoglobulin A gel at ambient temperature, revealing a velocity change of
1.5 m/s, whereas the attenuation change was $\approx$\SI{2}{\per\metre} throughout the
experiment~\cite{yuno-ohtaCharacterizationVLactoglobulinGelation2007}. These changes were very small in comparison with
heat-induced gelation, e.g., a 10 m/s decrease in velocity and \SI{10}{\per\metre}
increase in attenuation upon heating~\cite{corredigHeatInducedChangesUltrasonic2004}. Thus, our previous and current
results show that changes in the ultrasonic parameters in the SC system
(with or without OVA) during acidification are induced with only a small
change
caprate–induced gelation and with relatively large attenuation changes
despite the lack of heat treatment~\cite{yuno-ohtaCharacterizationVLactoglobulinGelation2007,corredigHeatInducedChangesUltrasonic2004}.

\subsection{Microstructure during gelation as assessed with CLSM imaging}
\begin{figure*}
\includegraphics[width=\textwidth]{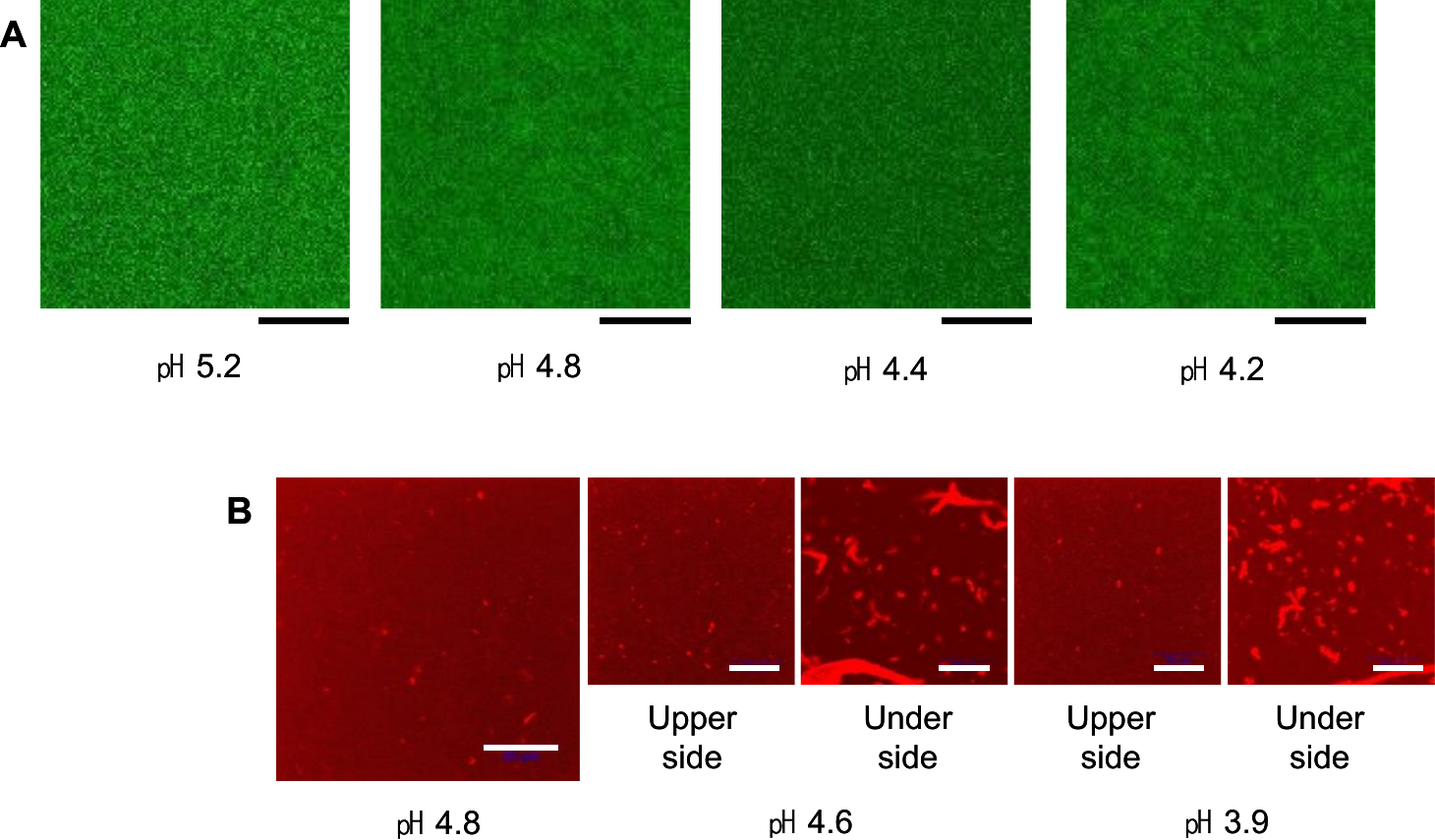}
\caption{CLSM image changes in SC or OVA solely as decreasing of pH.
A, SC; B, OVA; scale bar indicates \SI{5}{\micro\metre}.}\label{fig:CLSM_sole}
\end{figure*}
Figure~\ref{fig:CLSM_sole} presents data for the changes in images of SC or OVA protein
that coincided with the temporal decrease in pH owing to the addition of
GDL. At pH 5.2, although the SC solution was still turbid, numerous small
and sphere-shaped particles were observed but no interacting particles
were observed among SC particles (Fig. 3A; upper images). As the pH
continued to decrease over time, the SC underwent gelation, and faint
amorphous structures were visible at pH 4.8 and pH 4.2. At these pH
values, samples were gelled. Few or no structures were visible at pH 4.4,
and no movement was observed, indicating that the sample had gelled.
Given that pH 4.4 is the closest pH to the pI of SC, the gel particles became
too small to be visible with CLSM. On the other hand, during acidification,
OVA itself precipitated without gelling (Fig. 3B, lower images). Notably,
CLSM revealed a continuous movement of the precipitates until gelation.

\begin{figure*}
\includegraphics[width=\textwidth]{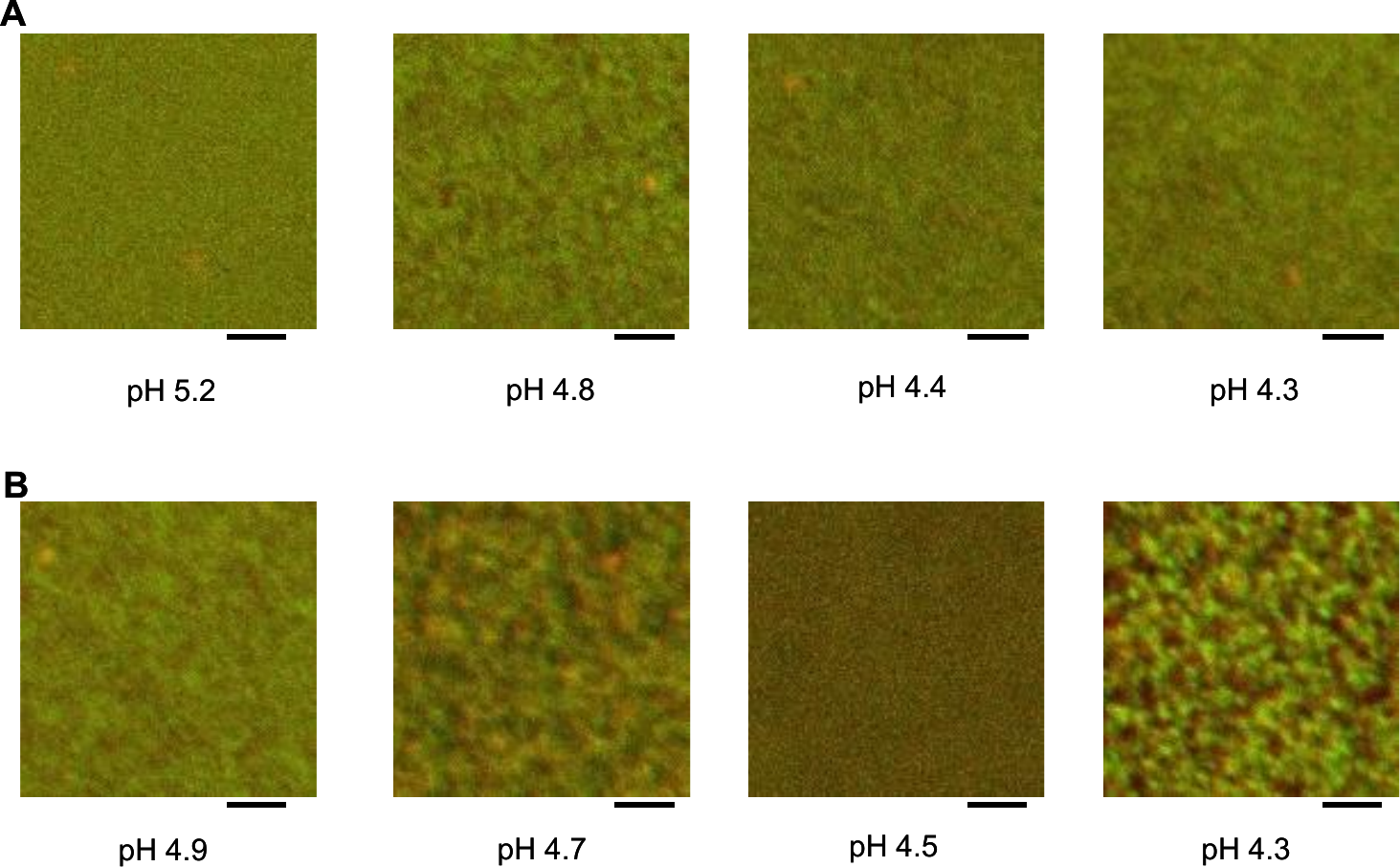}
\caption{CLSM image changes in SC and OVA mixed suspension as decreasing
of pH. A, 8\% SC+2\% OVA; B, 5\% SC+5\% OVA; scale bar indicates \SI{5}{\micro\metre}.}\label{fig:CLSM_mix}
\end{figure*}
Figure~\ref{fig:CLSM_mix} shows the acid-induced evolution of the structures of two mixed
systems: 8\% SC + 2\% OVA (Fig.~\ref{fig:CLSM_mix}A, upper images) and 5\% SC + 5\% OVA
(Fig.~\ref{fig:CLSM_mix}B, lower images). In general, the 5\% SC + 5\% OVA samples (Fig. 4B)
had coarser structures than the 8\% SC + 2\% OVA samples, the exceptions
being the pH 4.4 / 8\% SC + 2\% OVA sample and the pH 4.5 / 5\% SC + 5\%
OVA sample, as both of these pH values are close to the pI. Interestingly,
only very small particles were observed in the pH 4.5 / 5\% SC + 5\% OVA
sample, although the lack of movement hinted that the sample had gelled.
This result indicated that the particle size of the gels decreases
substantially at pH values near the pI, even in mixed gels.
Notably, both constituents were well mixed within optical resolution, as
the network appeared uniformly yellow-green (equal parts green for SC and
red for OVA), although several red (OVA) aggregates were observed (Fig.
4A, pH 4.8, 4.4, 4.3; Fig. 4B, pH 4.7, 4.3).

\begin{figure}
\includegraphics[width=\columnwidth]{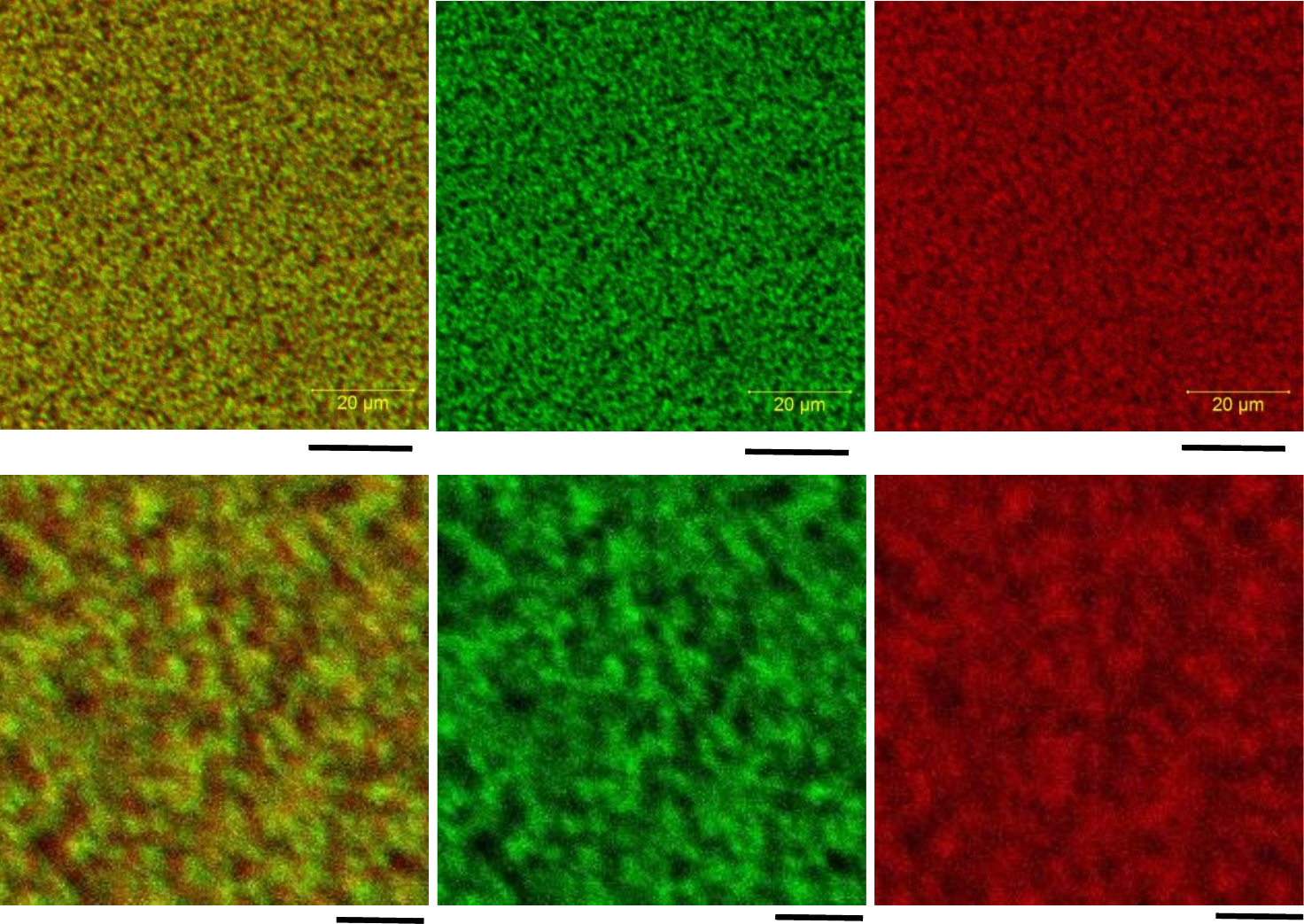}
\caption{CLSM image of acidified 5\% SC+ 5\% OVA mixed gel. The left CSLM
image show the overlay of the two separately labeled SC and OVA
proteins. The images in the middle show SC fraction covalently
labeled with DyLight 408 and presented with green filter and the
right image shows OVA fraction covalently labeled with DyLight
633 and presented with red filter. bars of upper and lower sides
indicate \SI{20}{\micro\metre} and \SI{5}{\micro\metre}, respectively.}\label{fig:CLSM_acid}
\end{figure}

Figure~\ref{fig:CLSM_acid} shows separate and merged images for the respective proteins in
the 5\% SC + 5\% OVA gel at pH 4.2. The upper and lower images were
acquired at different magnifications. Indeed, the red and green channels
are superimposable, indicating that SC and OVA acted cooperatively
during gel formation (upper images). However, the OVA aggregates were
spread out a little broader around the SC network (lower images; also see
Fig. S1, S2, S3 and S4).

\subsection{Ultrastructure of the SC and OVA mixed-protein gel}
\begin{figure*}
\includegraphics[width=\textwidth]{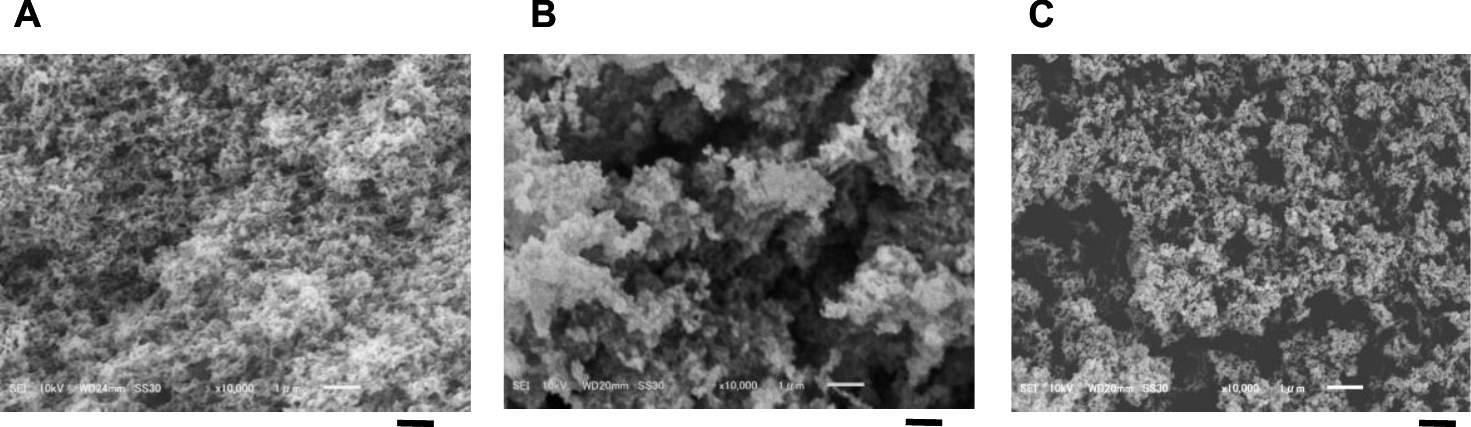}
\caption{SEM of the gels at pH 4.4: A, 10\% SC; B, 8\% SC+2\% OVA; c, 5\% SC+5\% OVA
gels. Scale bar indicates \SI{1}{\micro\metre}.}\label{fig:SEM}
\end{figure*}

As mentioned above, we found that the SC system and the SC + OVA
mixed-protein system at pH values near the pI had smaller particles than
the systems at pH values above or below the pI. We therefore performed
SEM to determine any differences in the network at pH values near the pI
($\approx$4.5) with the increase in OVA ratio. SEM revealed that these gels had
distinct ultrastructures at pH values near the pI (Fig.~\ref{fig:SEM}). The SC gel had a
homogeneous structure with small pores, whereas the mixed gels with OVA
had distinct microstructures with larger pores. In particular, the 8\% SC +
2\% OVA sample had larger aggregates, in agreement with the CLSM
images that showed that the 8\% SC + 2\% OVA gel had larger particles than
the 10\% SC and 5\% SC + 5\% OVA gels.

\section{Discussion}
This study investigated the interaction between SC and OVA during
GDL-mediated gelation using complementary instrumental analyses.
CLSM and SEM revealed that both proteins were well mixed and
cooperatively participated in gel formation. We suggest that SC and
OVA can collaborate during gelation to make a well-developed
three-dimensional structure when mixed at a 1:1 molar ratio, but it is
likely that SC is mainly responsible for gel-network formation.
High-resolution ultrasonic spectroscopy is a nondestructive means of
elucidating structural changes that occur within proteins during
acidification-induced gelation and can also detect protein aggregation
during this process~\cite{buckinUltrasonicShearWave2001}. Furthermore, ultrasound spectroscopy is very
sensitive and therefore can be used to help us understand the phase
transitions within proteins, e.g., the transition~\cite{powrie1986chemistry} of a protein from the
soluble to gelled state, and this technique can be combined with other
rheological analyses.
Kudryashov et al.~\cite{kudryashovUltrasonicHighResolutionLongitudinal2000} investigated the process by which milk gels
form during acidification with GDL. They used 2\% $\beta$-LG or 3.3\% milk
protein (fresh milk) and 2\% GDL, which is a higher proportion of GDL
than that used in our experimental conditions. They found that the
ultrasonic parameters changed along with the temporal decrease in pH
from 5.6 to 4.6 owing to solubilization of colloidal calcium phosphate. In
other work, Corredig et al.~\cite{corredigApplicationUltrasonicSpectroscopy2004} compared ultrasonic properties during
the gel-formation process using different methods of gelling fresh milk
and commercial milk whey (renneting milk, acidifying milk, and
heat-gelling whey protein). They found that formation of a gel in the
acidified milk system caused very small changes in ultrasonic velocity
and attenuation. They mentioned about the reason that the changes in
the structures of casein micelles may occur in the system prior to
gelation. Several studies have produced conflicting results. For example,
McClements\cite{mcclementsPrinciplesUltrasonicDroplet1996,mcclementsUltrasonicMeasurementsParticle2006} stated that changes in attenuation can arise from
changes in particle size, whereas Dalgleish et al.~\cite{dalgleishStudiesAcidGelation2004} reported that
although the attenuation increased at the early stage of acidification,
the particle size decreased in direct measurements using diffusing wave
spectroscopy.
Our system differed from the systems used in those studies. For
example, our material (SC) contained approximately one-tenth part
minerals (Ca and P, etc.) of fresh milk. The velocity decreased and the
attenuation increased in all three systems along with the decrease of pH,
which generally reflected the trend of gel formation in these three
systems, i.e., 10\% SC, 8\% SC + 2\% OVA, and 5\% SC + 5\% OVA. Due to
low mineral content of SC, it seems that colloidal calcium phosphate
may not have had any effect at all on gelation in our experimental
system. It seems that differences in components may be an important
factor for the observed differences of ultrasound property changes
between the Dalgleish et al. study (using fresh milk) and our present
study (using SC).
Our current study suggests that, although the differences in
ultrasound properties between the 10\% SC gel and the 8\% SC + 2\% OVA
mixed-protein gel
were
not
clear,
the
dynamic
viscoelastic
measurements revealed differences in their dynamic modulus—in other
words, the gel network structure changed depending on the SC:OVA
ratio. Furthermore, our CLSM results indicated that the observed
differences in rheological properties are attributable to differences in gel
networks in the mixed-protein systems. In those mixed systems, both
proteins were well mixed at pH values near their pI, probably owing to
loss of electrostatic repulsion. The pI for each of SC or OVA is $\approx$4.5, and
thus it remains uncertain why the pH of samples containing more OVA
decreased
supplemental figure S5). However, it is very interesting that the
mixed-protein systems formed gels although phase separation during
acidification
microstructures of 8\% SC + 2\% OVA and 5\% SC + 5\% OVA revealed that
these proteins respond differently to acidification by GDL.
more
rapidly
occurred
than
only
for
those
10\%
containing
OVA.
more
Comparison
SC
of
(see
the
Vasbinder et al.~\cite{vasbinderGelationCaseinWheyProtein2004} investigated the role of whey in reconstituted skim
milk, and the network that was formed was analyzed with CLSM via
separate staining for casein and whey protein isolate (WPI). They found
that whey protein aggregates were not linked to the casein micelles and
formed without phase separation when mixed with casein. The WPI
aggregates could interact with the casein fraction when the pH
approached the pI values for WPI and casein, and the casein and whey
formed a gel shortly thereafter. We conclude that the inclusion of OVA
(like WPI
three-dimensional network of the resulting gel depending on the
proportion of OVA.
~\cite{vasbinderGelationCaseinWheyProtein2004}) in a
SC-based gel
modifies the texture
and
Whereas there are many experimental techniques for understanding
the texture of biomolecular gels, the combined use of CLSM and SEM
can help us visualize how the different components of a mixed-protein
system contribute to the gel microstructure.

\section{Conclusion}
We investigated the gelation and resulting gel network of a
mixed-protein system derived from SC and OVA in the presence of GDL
to determine how OVA affects the gelation process and the gel network.
The results indicated that increased OVA content lessened the
mechanical strength of the gel yet accelerated gelation. Although CLSM
and SEM revealed that the SC and OVA solutions mixed uniformly, SC
likely led to the formation of the gel network and governed the porosity
of the resulting network depending on the SC:OVA ratio. Based on our
results, it appears that the inclusion of OVA in a SC-based gel modifies
the texture and three-dimensional network of the resulting gel
depending on the proportion of OVA. Utilization of CLSM for analysis of
a mixed-protein gel is useful for determining differences in the
networks of mixed-protein gels and the contribution of each protein to
gel formation.

\begin{acknowledgments}
The authors would like to thank Meiji Co., Ltd. (Kanagawa, Japan) for
providing sodium caseinate samples. This work was supported in part
by a Grant-in-Aid for Scientific Research on Priority Areas (No. 16K00838) from the Ministry of Education, Culture, Sports, Science and Technology in Japan.
\end{acknowledgments}

\bibliography{ohta}
\end{document}